\documentclass[aps,pre,twocolumn,showpacs,amsmath,amssymb,superscriptaddress]{revtex4}

\usepackage{graphicx}
\usepackage{dcolumn}
\usepackage{bm}

\begin{document}

\title{A Weakly Nonlinear Analysis of Impulsively--Forced Faraday Waves}

\author{Anne Catll\'{a}}
\email{a-catlla@northwestern.edu} \affiliation{Department of 
Engineering Sciences and Applied Mathematics, Northwestern
University, Evanston, IL 60208}
\author{Jeff Porter}
\affiliation{Instituto Pluridisciplinar, Universidad Complutense de Madrid,
28040 Madrid, Spain}
\author{Mary Silber}
\affiliation{Department of Engineering Sciences and Applied
Mathematics, Northwestern University, Evanston, IL 60208}

\date{\today}

\begin{abstract}
  
  Parametrically-excited surface waves, forced by a repeating sequence
  of $N$ delta-function impulses, are considered within the framework
  of the Zhang-Vi\~nals model~\cite{ZV97}.  The exact
  impulsive-forcing results, in the linear and weakly nonlinear
  regimes, are compared with results obtained numerically for
  sinusoidal and multifrequency forcing. We find surprisingly good
  agreement between impulsive forcing results and those obtained using
  a two-term truncated Fourier series representation of the impulsive
  forcing function.  With impulsive forcing, the linear stability analysis 
  can be carried out exactly and leads to an implicit equation for the neutral 
  stability curve. As noted previously by Bechhoefer and Johnson~\cite{BJ96}, 
  in the simplest case of $N=2$ {\it equally-spaced} impulses per period
  (which alternate up and down) there are only subharmonic modes of
  instability. The familiar situation of alternating subharmonic and
  harmonic resonance tongues emerges only if an asymmetry in the
  spacing between the impulses is introduced.  We extend the linear
  analysis for $N=2$ impulses per period to the weakly nonlinear
  regime, where we determine the leading order nonlinear saturation of
  one-dimensional standing waves as a function of forcing strength.
  Specifically, an analytic expression for the cubic Landau
  coefficient in the bifurcation equation is derived as a function of
  the dimensionless spacing between the two impulses and the fluid
  parameters that appear in the Zhang-Vi\~nals model.  As the
  capillary parameter is varied, one finds a parameter region of
  wave amplitude suppression, which is due to a
  familiar 1:2 spatio-temporal resonance between the subharmonic mode
  of instability and a damped harmonic mode. This resonance 
  occurs for impulsive forcing even when harmonic resonance tongues
  are absent from the neutral stability curve. The strength of this
  resonance feature can be tuned by varying the spacing between the
  impulses.  This finding is interpreted in terms of a recent
  symmetry-based analysis of multifrequency forced Faraday
  waves~\cite{PTS04,TPS04}.
\end{abstract}


\maketitle

\section{I. Introduction}

Standing waves form spontaneously on the free surface of a fluid layer
when it is subjected to a time-periodic vertical vibration of
sufficient strength. The waves, which result from a symmetry-breaking
parametric instability, organize themselves into remarkably regular
patterns, as first described by Faraday in 1831~\cite{F1831}.  In
modern experiments, a variety of container geometries, fluids and
forcing functions have been employed, resulting in a rich variety of
standing wave patterns, as well as a range of dynamical responses.
See~\cite{MH90,MFP98} for reviews on Faraday waves.  In this paper,
we investigate the response of the Faraday system to an idealized
forcing function consisting of a periodic sequence of delta-function
impulses. The results are contrasted with those obtained in the
classical cases of single and multifrequency forcing.

The versatility of the Faraday system for investigating pattern
formation owes much to the vastness of its control parameter space.
By varying the forcing frequency and amplitude, as well as the fluid
properties, experimentalists employing a  sinusoidal
acceleration function have teased out spatially regular patterns
ranging from stripes, hexagons and squares to targets, spirals and
quasipatterns~\cite{CAL92,KG96,BV97,WMK00}.  The seminal experiments
of Edwards and Fauve~\cite{EF94} showed how the addition of a second
commensurate frequency component in the forcing function could lead to
even greater versatility of this system, via the controlled
introduction of additional length scales. Two-frequency forcing has
led to a variety of exotic superlattice patterns~\cite{KPG98,AF98}, as
well as quasipatterns~\cite{EF93,AF00a}, triangular
patterns~\cite{M93} and localized structures~\cite{AF00b}. A detailed
description of patterns readily achieved in two-frequency Faraday
experiments can be found in~\cite{AF02}. Recent experiments by
Epstein and Fineberg~\cite{EF04} have shown how a third perturbing
frequency can be used to rapidly switch between the novel patterns
achieved with two-frequency forcing.

Theoretically,  the Faraday system presents a number of
challenges due both to the explicit time-dependence of the forcing and
the free-boundary nature of the problem; for a partial review of
theoretical work see~\cite{MFP98}.  Consequently, models incorporating
simplifying assumptions have often been relied on, in conjunction with
numerical linear stability analysis or perturbative methods.  In this
manner, much has been accomplished.  Linear stability results build on
the classic paper of Benjamin and Ursell~\cite{BU54}, who explained
Faraday's observation of a subharmonic response of the fluid layer to
sinusoidal forcing. Specifically, they showed that the linear
stability problem, in the case of an ideal fluid, reduces to a Mathieu
equation, for which the natural frequency of oscillation is given by
the inviscid dispersion relation.  Subsequent investigators carried out
linear stability analyses on the full fluid problem with sinusoidal
forcing~\cite{KT94,BF95,K96,CV97}. These results were also extended to
two-frequency forcing in~\cite{BET96}. Theoretical understanding of the
nonlinear problem progressed with the introduction of a quasipotential
formulation of the Faraday wave problem by Zhang and
Vi\~{n}als~\cite{ZV97}. This model, which describes small amplitude
surface waves on a semi-infinite, weakly viscous fluid layer, eliminates
the bulk flow, assumed to be potential, from the description. The
free boundary $z=h({\bf x},t)$ $({\bf x}\in\mathbb{R}^2)$ is then
prescribed by an evolution equation for $h$, which is coupled to an
evolution equation for a surface velocity potential $\Phi$; see
Section~\ref{ref:ZVmodel}.  Zhang and Vi\~nals used this model to
further investigate nonlinear effects via an asymptotic expansion,
which demonstrated the importance of resonant triads (three-wave
resonance) in the pattern selection process.  Later, Chen and
Vi\~{n}als~\cite{CV99} used perturbation methods to derive amplitude
equations directly from the Navier-Stokes equations assuming only that
the depth of the fluid layer is (effectively) infinite.  These
approaches yield good agreement with experiments in the relevant
parameter regimes, but typically require a numerical evaluation of
relevant quantities to make the comparison.

An alternative theoretical approach to the Faraday wave pattern
selection problem is based in equivariant bifurcation
theory~\cite{GSS88,CK91}. This ``model-independent'' treatment of the
symmetry-breaking bifurcations relevant to the Faraday experiment has
proved especially fruitful in analyzing spatially-periodic
superlattice patterns~\cite{SP98,TRHS00}. An extension of this
approach, tailored to the multifrequency Faraday problem, uses
spatio-temporal symmetry arguments to derive amplitude equations based
on the patterns' symmetries. The form of the coefficients in the
equations reflect the temporal (parameter) symmetries of the forcing
function and the time-reversal symmetry that is broken by the damping.  
The equations describe not only the critical modes, but
also certain damped modes that are nonlinearly driven by them. These
studies reveal the role of the various forcing parameters 
(frequencies, amplitudes, and relative phases) in nonlinearly coupling
the driven modes to key weakly-damped modes.  The results give insight 
into the pattern selection process and can be used to explain
a number of characteristics of the experimentally observed
patterns~\cite{SS99,TS02,PS04,PTS04,TPS04}.

The investigations reviewed above focus on Faraday waves
parametrically-excited by sinusoidal and multifrequency forcing
functions. An alternative forcing function, consisting of a periodic
sequence of delta-function impulses, was proposed by Bechhoefer and
Johnson~\cite{BJ96} who indicated how the linear analysis could be
greatly simplified in this case. (This idealization of parametrically
forced systems has been put forth in a variety of physical contexts;
see, for example, \cite{H72,HC73,PVS99}.)  Recently, Huepe and
Silber~\cite{HSprep} showed that the linear analysis of the
impulsively-forced Faraday problem for a viscous fluid, as proposed
in~\cite{BJ96}, breaks down in certain parameter regimes. In these
regimes, the flow in the fluid bulk cannot be (instantaneously)
matched across the delta impulses. However, these complications are
absent from the Zhang-Vi\~nals formulation, which describes the
evolution of the free surface, with the bulk flow eliminated from the
model. In this paper we extend the results of Bechhoefer and Johnson
on impulsively-forced Faraday waves into the weakly nonlinear regime,
within the framework of the Zhang-Vi\~nals model. 

Following the method of~\cite{H72}, we develop a simple modular
approach to constructing the stroboscopic map associated with the
linear stability problem. Specifically, we construct the
linear maps which evolve perturbations of wavenumber $k$ from one
impulse in the sequence to the next. The linear map associated with
one period of the forcing is then obtained by multiplication of $N$
matrices, one for each impulse in the periodic sequence.  From this
simple construction we can derive explicit expressions for the
Floquet multipliers associated with the linear stability problem, and
arrive at an equation (implicit if $N>2$) that describes the neutral
stability curve. We analyse the results in the simplest case of $N=2$
impulses in detail, and compare them with those obtained with an
$M$-term truncated Fourier series representation of the impulsive
forcing function. While the neutral stability curves are poorly
approximated by the truncated Fourier series even for very large $M$, 
the onset parameters of critical forcing strength and wavenumber 
are quite accurately determined with a single-term sinusoidal 
approximation ({\it i.e.}, for $M=1$).

We extend the analysis for $N=2$ impulses per period to the weakly
nonlinear regime for one-dimensional surface waves.  A key observation
of Bechhoefer and Johnson~\cite{BJ96} is that there are no harmonic
resonance tongues for a sequence of $N=2$ impulses if they are equally
spaced in time ({\it i.e.}, an up impulse and then a down impulse half
a period later).  We show that, despite the absence of harmonic
instabilities, there can still be a resonant interaction between the
excited subharmonic and damped harmonic modes which leads, in one
dimension, to an associated degradation in the nonlinear response of
the fluid to the vibration in the weakly nonlinear regime. This is
similar to the spatio-temporal 1:2 resonance that occurs in the case
of sinusoidal forcing~\cite{ZV97}, for which subharmonic and harmonic
instabilities alternate with increasing perturbation wavenumber
$k$. Thus, although the harmonic modes can never be linearly excited
in this particular impulsive case, these modes can nonetheless be driven 
nonlinearly by the critical subharmonic modes and in this way influence 
the nonlinear evolution of the instability.  In addition to analysing the 
case of $N=2$ {\it equally-spaced} impulses, we explore the effects 
of varying the spacing between the impulses so that they are no longer 
exactly half a period apart.  We find that this asymmetry in impulse timing
enhances or diminishes the influence of the 1:2 resonance on the
standing wave amplitude depending on the sense in which it is applied.
We understand this result by considering a two-term Fourier series
approximation to the impulsive forcing function, and applying recent
results pertaining to multifrequency forcing~\cite{PTS04,TPS04}.

Our paper is organized as follows. In the next section we present the
Zhang-Vi\~nals model of Faraday waves, and then carry out the linear
stability analysis in the general case of $N$ impulses per forcing period.
We then focus on the simplest case of $N=2$ and examine in detail 
the sequence of subharmonic and harmonic resonance tongues as a
function of the dimensionless spacing between the impulses.  We
compare our results with those obtained with sinusoidal and
multifrequency forcing functions within the context of a truncated
Fourier series representation of the impulsive forcing function.  In
section III we derive the cubic bifurcation equation that determines
the amplitude of one-dimensional spatially-periodic surface waves
driven by impulsive forcing, focusing on the 1:2 spatio-temporal
resonance feature. We compare our weakly nonlinear results with those
obtained with single and two-frequency forced Faraday waves.  Finally,
in Section IV, we summarize our results and indicate some promising
directions for subsequent investigations.

\section{Linear results for the Zhang-Vi\~nals Faraday wave model}
\label{ref:ZVmodel}

This section introduces the Zhang-Vi\~{n}als model for Faraday waves.
A linear stability analysis is then performed for waves excited by a
periodic sequence of $N$ impulses and detailed results are presented
for the simplest case of $N=2$.

\subsection{Zhang-Vi\~{n}als model}

The quasipotential formulation of the Faraday wave problem, due to
Zhang and Vi\~{n}als~\cite{ZV97}, is derived from the Navier-Stokes
equations assuming small amplitude surface waves on a deep,
nearly inviscid fluid layer.  It describes the free surface height
$h(\textbf{x},t)$ and surface velocity potential $\Phi(\textbf{x},t)$
of a fluid subjected to a (dimensionless) periodic vertical
acceleration function $G(t)$.  Employing these equations greatly
simplifies our calculations since the flow in the bulk does not appear
explicitly --- we need only track the behavior of the free surface, a
function of the horizontal coordinate $\textbf{x}\in\mathbb{R}^2$ and time
$t$.
The Zhang-Vi\~{n}als model takes the form
\begin{subequations}
\begin{align}
(\partial_t - \gamma \nabla^2)h - \hat{D}\Phi &= N_{1}(h,\Phi), 
\label{eq:zva}\\
(\partial_t-\gamma \nabla^2) \Phi-(\Gamma_{0}\nabla^2-G_{0}+G(t))h
&= N_{2}(h,\Phi) \label{eq:zvb},
\end{align}\label{eq:zv}
\end{subequations}
where the nonlinear terms in (\ref{eq:zv}) are given by
\begin{subequations}\label{eq:zvnlin}
\begin{align}
N_{1}(h,\Phi) = &\;-\nabla\cdot(h\nabla\Phi)
+{1\over2}\nabla^2(h^2\hat{D}\Phi)-\hat{D}(h\hat{D}\Phi) \nonumber \\ 
& +\hat{D}(h\hat{D}(h\hat{D}\Phi)+{1\over2}h^2\nabla^2\Phi), \\
N_{2}(h,\Phi) = &\;\frac{1}{2}(\hat{D}\Phi)^2-\frac{1}{2}(\nabla\Phi)^2 
-{1\over2}\Gamma_0\nabla\cdot((\nabla h)(\nabla h)^2) \nonumber \\ 
& -(\hat{D}\Phi)(h\nabla^2\Phi+\hat{D}(h\hat{D}\Phi)).
\end{align} 
\end{subequations}
Here the operator $\hat{D}$ multiplies each Fourier component by
its wavenumber, e.g.
$\hat{D}e^{i\textbf{k}\cdot\textbf{x}}=|\textbf{k}|e^{i\textbf{k}\cdot\textbf{x}}$,
and the dimensionless fluid parameters are
\begin{equation}
\gamma \equiv \frac{2 \nu k_0^2}{\omega}, \quad \Gamma_{0} \equiv
\frac{\Gamma k_0^3}{\rho \omega^2}, \quad G_0 \equiv \frac{g
k_0}{\omega^2}, \label{eq:params}
\end{equation}
where $g$ is the usual gravitational acceleration, $\omega$ is the
forcing frequency ($2\pi$ divided by the forcing period), $\nu$ is the 
kinematic viscosity, $\rho$ is the density, and $\Gamma$ is the surface 
tension.  The wavenumber $k_0$ is chosen to satisfy the inviscid 
dispersion relation
\begin{equation}
gk_0+ \frac{\Gamma k_0^3}{\rho}= \left(\frac{\omega}{2}\right)^2,
\label{eq:disp}
\end{equation}
where $\frac{\omega}{2}$ is the frequency associated with the typical 
subharmonic response of Faraday waves.  After dividing (\ref{eq:disp}) by 
$\omega^2$ we find $G_0+\Gamma_0 = \frac{1}{4}$.  $G(t)$ describes the 
applied acceleration; since time has been scaled by the forcing frequency 
$\omega$, the dimensionless period of $G(t)$ is $2\pi$.

\begin{figure}
\includegraphics[width=3.3in]{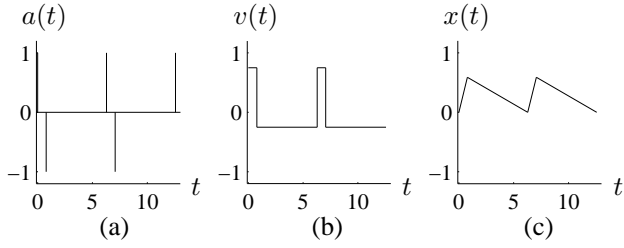}
\caption{Example of (a) acceleration, (b) velocity, and (c) position 
of the fluid container over time in the case of two impulses per period. 
The positions of the vertical lines in the acceleration function denote 
the locations of the delta functions; here they have equal magnitude.
\label{fig:example1}}
\end{figure}

We write the impulsive forcing function in the form
\begin{equation}
G_{imp}(t) = \sum_{n=-\infty}^{\infty}f_n\delta(t-t_n)\ .\label{eq:impacc}
\end{equation}
It is parameterized by the locations $t_n$ of the impulses
($t_n<t_{n+1}$, $t_{n+N}-t_n=2\pi$) and by the amplitudes $f_n$
($f_{n+N}=f_n$), where $N$ is the number of impulses in the repeating
sequence.  The dimensionless amplitude $f_n$ of an impulse is given by
\begin{equation}
f_{n} \equiv \frac{v_n k_0}{\omega},\label{eq:vjump}
\end{equation}
where $v_n$ is the jump in velocity at time $t_n$.  We must also require  
\begin{equation}
\sum_{n=1}^{N} f_n = 0, \label{eq:trans}
\end{equation}
to prevent a net translation of the container (for suitably chosen
initial velocity).  An example of an impulsive acceleration function,
along with the corresponding velocity and position functions, is shown
in Fig. \ref{fig:example1}.

\subsection{Linear stability analysis}
\label{ssec:linstabanal}

\subsubsection{Calculations}
\label{ssec:calculations}

Our linear stability calculations follow the method of~\cite{BJ96}. 
 From (\ref{eq:zv}), the linear stability problem takes the form
\begin{equation}
\left[(\partial_t-\gamma \nabla^2)^2 \!-\! \hat{D}(\Gamma_0 \nabla^2 
\!-\! (G_0 \!-\! G(t))) \right] h(x,t)=0.
\label{eq:pdiff}
\end{equation}
We then write $h({\bf x},t) = p_k(t)e^{i{\bf k}\cdot{\bf x}}+c.c.$,
where $c.c.$ denotes complex conjugate, and find that $p_k(t)$
satisfies
\begin{equation}\label{eq:peq}
p_k^{\prime\prime} + 2\gamma k^2 p_k^{\prime} + [\gamma^2 k^4
+\Gamma_0 k^3+(G_0 - G(t))k]p_k = 0,
\end{equation}
where the prime denotes differentiation with respect to $t$ and the
$k$ subscript emphasizes the dependence of the solution on the
perturbation wavenumber $k=|{\bf k}|$.  Hereafter we focus on $k\ne 0$; the
$k=0$ mode cannot be excited due to mass conservation. (This is
reflected in the governing equations~(\ref{eq:zvnlin}) through the
spatial derivatives in the nonlinear terms which ensure that the
$k=0$ mode decouples from the other modes.)

Between each impulse, (\ref{eq:peq}) is simply the equation for a
damped harmonic oscillator, with solution
\begin{equation}
p_{k}(t) = A_{k,n} e^{(-\gamma k^2+i \omega_k)(t-t_{n})} +
c.c.,\:\:\:\: t\in(t_n, t_{n+1}).\label{eq:p}
\end{equation}
Here $\omega_k$ is the natural frequency of a wave with wavenumber
$k$, which is determined by the dispersion relation
\begin{equation}
\omega_k^2 = \Gamma_0 k^3 + G_0 k.
\label{eq:dispersion}
\end{equation}
We demand that $p_k(t)$ be continuous across each impulse, {\it i.e.},
$p_{k}(t_n^+)=p_{k}(t_{n}^-)\equiv p_k(t_n)$, where
$p_{k}(t_n^{\pm})\equiv \lim_{t\rightarrow t_n^{\pm}}p_k(t)$.
Integrating (\ref{eq:peq}) across the $n^{th}$ impulse we obtain 
the following jump condition:
\begin{equation}
p_{k}^\prime(t_n^+) - p_{k}^\prime(t_n^-) = f_nk\,p_{k}(t_n).
\label{eq:jump}
\end{equation}
This condition, together with the continuity requirement, yields the following 
map from $A_{k,n}$ to $A_{k,n+1}$:
\begin{equation}
 \left( \begin{array}{c}
    A_{k,n+1}^r \\ A_{k,n+1}^i
\end{array} \right)
= e^{-\gamma k^2 d_n} M_{k,n} \left( \begin{array}{c}
    A_{k,n}^r \\ A_{k,n}^i
\end{array} \right), \label{eq:A}
\end{equation}
where
\begin{equation}
\label{eq:Mn} 
M_{k,n} \equiv  \left( \begin{array}{cc} 
    c_{k,n} & -s_{k,n} \\
    s_{k,n}-F_{k,n+1} c_{k,n} &
    c_{k,n}+F_{k,n+1} s_{k,n}
\end{array} \right).
\end{equation}
Here $A_{k,n}^r$ ($A_{k,n}^i$) is the real (imaginary) part of
$A_{k,n}$ and $c_{k,n} \equiv \cos(\omega_k d_n)$, $s_{k,n} \equiv
\sin(\omega_k d_n)$, $F_{k,n} \equiv \frac{f_n k}{\omega_k}$. Note
that (\ref{eq:Mn}) depends on only two parameters: $d_n \equiv
t_{n+1}-t_n$, the time between the $n^{th}$ and $(n+1)^{st}$ impulses,
and $f_{n+1}$, the strength of the $(n+1)^{st}$ impulse.  

Piecing together the relationships between $A_n$ and $A_{n+1}$, $A_{n+1}$ and 
$A_{n+2}$, etc. across one period, we find the stroboscopic map that 
relates the solution of the linearized problem at time $t_n$ to the 
solution one period later ({\it i.e.}, after the sequence of $N$ impulses):
\begin{equation} \left( \begin{array}{c}
    A_{k,n+N}^r \\ A_{k,n+N}^i
\end{array} \right) = e^{-2\pi \gamma k^2} M_k \left( \begin{array}{c}
    A_{k,n}^r \\ A_{k,n}^i
\end{array} \right).\label{eq:AnN}
\end{equation}
Here $M_k \equiv M_{k,n+N-1}\cdots M_{k,n}$, where $M_{k,j}$ is
given by~(\ref{eq:Mn}).

The eigenvalues of $e^{-2\pi\gamma k^2}M_k$ determine the linear 
stability of the interface to disturbances of wavenumber $k$.  Note 
that $M_k$ has determinant one, so these eigenvalues ($\lambda_\pm$) 
can be related to the trace of $M_k$ by $e^{2\pi\gamma k^2}\lambda_\pm  
= \frac{1}{2}{\rm Tr}(M_k)\pm\sqrt{(\frac{1}{2}{\rm Tr}(M_k))^2-1}$. An 
instability exists when both eigenvalues (Floquet multipliers) are real 
and the magnitude of one exceeds unity.  The threshold condition is 
therefore
\begin{equation}\label{eq:stab}
\frac{1}{2}{\rm Tr}(M_k) = \pm \cosh(2\pi\gamma k^2),
\end{equation}
where `+' corresponds to the harmonic case (Floquet multiplier $+1$), 
and `$-$' corresponds to the subharmonic case (Floquet multiplier $-1$).  
This threshold condition defines an implicit equation for the neutral 
stability curve $f(k)$, where $f$ is an appropriate measure of the 
overall impulse strength (e.g., for $N=2$ we use $f=|f_n|=|f_{n+1}|$).

\begin{figure}[tb]
\includegraphics{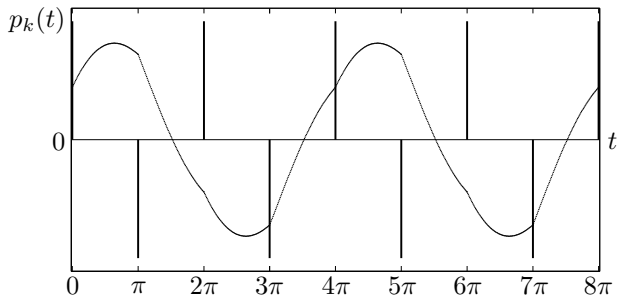}
\caption{Eigenfunction $p_k(t)$ at $k=k_c$, $f=f_c$ for an impulsive forcing
function with equally-spaced impulses.  The dimensionless parameters
used are $\gamma=0.17$, $\Gamma_0=0.19$ (roughly corresponding to
physical parameters $\nu = 0.20$ ${\rm cm^2/s}$, $\Gamma = 16$ dyn/cm,
$\rho = 1$ ${\rm g/cm}^3$, and $\frac{\omega}{2\pi} = 80$ H).  Note
the discontinuities in the derivative of $p_k(t)$ induced by the
impulses at $t=n\pi$.
\label{fig:eigenftn}}
\end{figure}

Examples of neutral stability curves for various acceleration
functions are presented in 
Figs.~\ref{fig:linear},\ref{fig:NscTruncEq}--\ref{fig:overlay}.  For
values of $f$ below the minimum of these curves, the trivial solution
($h=\Phi=0$) is stable.  When $f$ is increased above the critical
forcing amplitude $f_c$ at the curves' minimum (with corresponding
critical wavenumber $k_c$), the flat surface solution becomes
unstable to standing waves. An example of the associated critical
eigenfunction, $p_k(t)$ ($k=k_c$ and $f=f_c$), is shown in
Fig.~\ref{fig:eigenftn}.  Note the presence of kinks in $p_k(t)$ at
the impulses, a consequence of the jump condition~(\ref{eq:jump}).

\subsubsection{Linear stability results for $N=2$}

We now present a detailed analysis for the case of two impulses per period.  
In this illustrative example, the forcing function is parameterized by two
constants: $f$, which characterizes the magnitude of the velocity jump 
associated with each impulse (see~(\ref{eq:vjump})), and 
\begin{equation}
\Delta\equiv \frac{d_0-d_1}{2\pi}\in(-1,1)\ ,
\label{eq:deltadef}
\end{equation}
which measures the deviation from equal spacing of the impulses. Thus
$\Delta=0$ corresponds to equally-spaced impulses, and as $|\Delta|$
increases the spacing between the impulses becomes increasingly 
asymmetric; see Fig.~\ref{fig:example1} for an example where $\Delta<0$. 
Specifically, we set $f_n=(-1)^nf$, and
\begin{equation}
t_n=\pi\Bigl(n+(1-(-1)^n)\frac{\Delta}{2}\Bigr)
\label{eq:deltadef2}
\end{equation}
\noindent 
in~(\ref{eq:impacc}), where we have assumed, without loss of generality, 
that $t_0=0$.

It follows from~(\ref{eq:AnN}) that $A_{k,n+2}$ and $A_{k,n}$ are
related by
\begin{widetext}
\begin{eqnarray}
 \left( \begin{array}{c}
    A_{k,n+2}^r \\ A_{k,n+2}^i
\end{array} \right)
= e^{-2\pi\gamma k^2 } \left( \begin{array}{cc}
    \cos(2\pi\omega_k) - F_nc_ns_{n+1} & -\sin(2\pi\omega_k)+F_ns_ns_{n+1} \\
    \sin(2\pi\omega_k) + F_n s_{n+1}(s_n+F_nc_n) \quad & \quad
    \cos(2\pi\omega_k)+F_ns_{n+1}(c_n-F_ns_n)
\end{array} \right)
\left( \begin{array}{c}
    A_{k,n}^r \\ A_{k,n}^i
\end{array} \right),
 \label{eq:MTwo}\end{eqnarray}
where
\begin{equation}
c_n \equiv \cos(\omega_k d_n), \ s_n\equiv \sin(\omega_k d_n), 
\ d_n=\pi(1+(-1)^n\Delta), \ F_n = \frac{f_nk}{\omega_k}\ .
\end{equation}
\end{widetext}
Solving (\ref{eq:stab}) for the neutral stability curve $f(k)$, we
find
\begin{equation} f(k) =
\frac{2\omega_k}{k}\sqrt{\frac{\pm \cosh(2\pi\gamma
k^2)-\cos(2\pi\omega_k)}{\cos(2\pi\omega_k)-\cos(2\pi\omega_k\Delta)}}, 
\label{eq:asymlin}
\end{equation}
where $\omega_k$ is given by~(\ref{eq:dispersion}).
As before, `+' corresponds to the harmonic case and `--' to the
subharmonic case.  Note that $f(k)$ is invariant under $\Delta\to
-\Delta$, so we can restrict to $\Delta\in[0,1)$ for 
the remainder of the linear analysis. This is not, however, a symmetry
of the full problem and we will see that the sign of $\Delta$ affects
the nonlinear response of the fluid to the impulsive forcing function.

\begin{figure}[tb]
\includegraphics{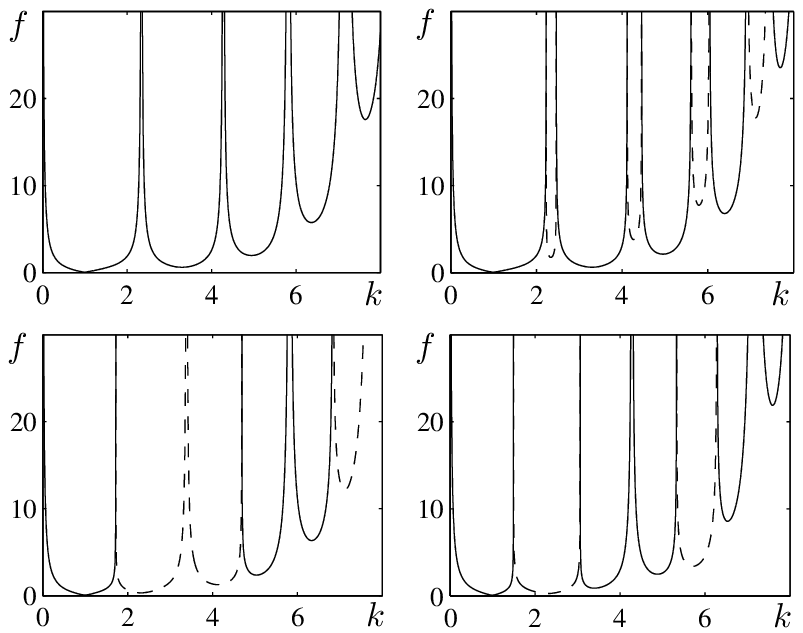}
\caption{Examples of neutral stability curves $f(k)$
from~(\ref{eq:asymlin}) for various two-impulse forcing functions.
Solid curves correspond to subharmonic tongues (Floquet
multiplier $-1$), while dashed curves indicate harmonic tongues
(Floquet multiplier $+1$). The spacing of the impulses, given 
by~(\ref{eq:deltadef}),
is (a) $\Delta = 0$, (b) $\Delta = 0.05$, (c) $\Delta = \frac{1}{3}$,
(d) $\Delta = \frac{1}{2}$.  
The dimensionless parameters in the calculations
are $\gamma=0.02$ and $\Gamma_0=0.04$ (corresponding to fluid parameters
$\nu = 0.1$ ${\rm cm^2/s}$, $\Gamma = 16$ dyn/cm, $\rho = 1$ ${\rm g/cm}^3$,
and $\frac{\omega}{2\pi} = 20$ Hz). \label{fig:linear}}
\end{figure}

\begin{figure}[tb]
\includegraphics{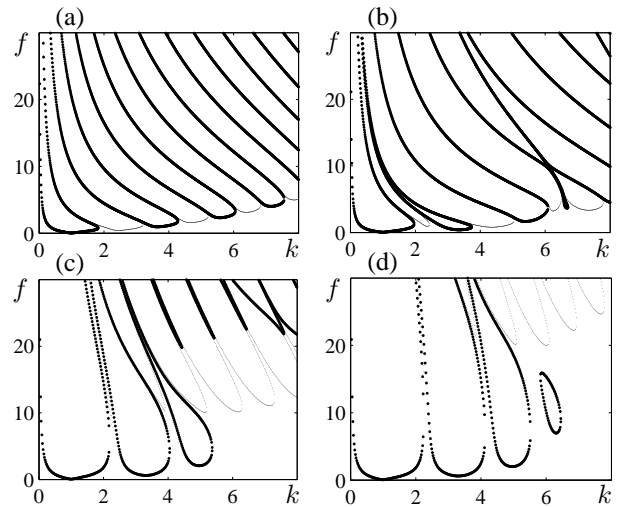}
\caption{Neutral stability curves for truncated Fourier series
approximations of equally spaced impulses with (a) 1 term, (b) 2
terms, (c) 10 terms, and (d) 20 terms. Large points correspond to 
subharmonic tongues and the smaller points to harmonic tongues.
Parameters are the same as those in Fig. \ref{fig:linear}.  Curves
were generated using the method described in~\cite{KT94,BET96}.
\label{fig:NscTruncEq}}
\end{figure}

The expression for the neutral stability curve $f(k)$, given
by~(\ref{eq:asymlin}), diverges at each value of $k$ for which
$\cos(2\pi\omega_k)=\cos(2\pi\omega_k\Delta)$. This leads to the
structure of clearly demarcated resonance tongues seen in
Fig.~\ref{fig:linear}.  The presence of vertical boundaries
between resonance tongues contrasts sharply with the classical 
case of sinusoidal forcing (Fig.~\ref{fig:NscTruncEq}(a)), where
multiple instabilities (involving distinct tongues) can occur  
for a single wavenumber.   We now examine in more detail the 
spacing and sequence of resonant tongues in the case of two 
impulses per period.  It proves useful, for the purposes of this 
analysis, to use $\omega_k$ as the independent variable, where 
$\omega_k$ and $k$ are related through the monotonic dispersion 
relation~(\ref{eq:dispersion}).

The divergence of the neutral curves, which leads to vertical
asymptotes separating the resonance tongues, is due to the denominator
in expression~(\ref{eq:asymlin}).  We therefore introduce
\begin{equation}
D(\omega_k)\equiv \cos(2\pi\omega_k)-\cos(2\pi\omega_k\Delta),
\label{eq:D}
\end{equation}
and denote the successive zeroes of $D(\omega_k)$ by $\omega_k^j$
($\omega_k^{j}<\omega_k^{j+1}$, $\omega_k^0=0$).  Since for $k>0$ 
the numerator of the expression inside the square root in~(\ref{eq:asymlin}) 
is strictly positive in the harmonic case and negative in the 
subharmonic case, it follows that $D(\omega_k)>0$ for harmonic resonance 
tongues and $D(\omega_k)<0$ for subharmonic resonance tongues.

In the general case, resonance tongues will alternate between harmonic 
and subharmonic as $k$ (equivalently, $\omega_k$) increases and $D(\omega_k)$ 
alternates in sign.  However, for special values of $\Delta$, successive 
subharmonic (or harmonic) tongues may occur.  Specifically, this happens 
for some $j$ if $D(\omega_k^j)=D^{\prime}(\omega_k^j)=0$.
We denote these double zeroes by $\hat{\omega}_k^m,\ m\in\mathbb{Z^+}$ 
($\hat{\omega}_k^{m}<\hat{\omega}_k^{m+1}$).  It is straightforward
to show that these degenerate values occur only when
$\cos(2\pi\hat{\omega}_k^m)=\cos(2\pi\hat{\omega}_k^m\Delta)=\pm
1$, which implies $\Delta=\frac{p}{q}$ for some co-prime integers
$p,q$ ($0 \leq p< q$).  There are thus two possibilities: either 
both $p$ and $q$ are odd, or one of them is even.  If they
are both odd, then $\hat{\omega}_k^m=\frac{m}{2}q$; successive
subharmonic (harmonic) tongues occur when $m$ is even (odd). (See
Fig.~\ref{fig:linear}(c), generated with $\Delta=\frac{1}{3}$.) 
In contrast, if either $p$ or $q$ is even (or $p=0$), then only 
subharmonic tongues can occur in succession; these are separated by vertical
asymptotes at $\hat{\omega}_k^m=mq$.  (See Fig.~\ref{fig:linear}(a) 
and (d) for examples of this case.)  Note that when $\Delta=0$ there 
are vertical asymptotes at each integer value of $\omega_k$ separating 
successive subharmonic tongues.  In other words, there are no harmonic 
resonance tongues for equally-spaced impulses (Fig.~\ref{fig:linear}(a)), 
as previously noted by Bechhoefer and Johnson~\cite{BJ96}.

The maximum possible width for the resonance tongues occurs when
$\Delta=0$, in which case $\omega_k^{j+1}-\omega_k^j=1$.  As
$\Delta$ is perturbed away from $0$, the vertical asymptotes
(located at $\omega_k^j=\hat{\omega}_k^j=j$ for $\Delta=0$) split, and 
a thin harmonic tongue appears between them thereby separating the 
two fat subharmonic tongues (see, for example, Fig.~\ref{fig:linear}(b)).  
Analogous results hold when $\Delta$ is perturbed away from other 
rational numbers.

\begin{figure}[tb]
\includegraphics{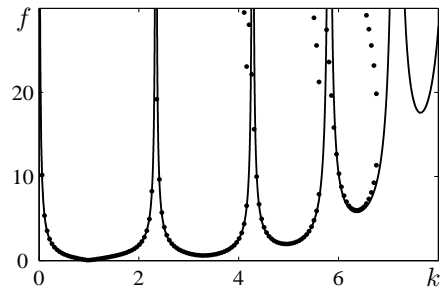}
\caption{Neutral stability curve for impulsive forcing with $\Delta=0$
(solid lines) and a truncated Fourier series approximation with 50 terms
(dots).  Note that the 5th (rightmost) resonance tongue is not
captured at all by the approximation. 
Parameters are as in Fig.~\ref{fig:linear}.
\label{fig:overlay}}
\end{figure}

\subsubsection{Comparison with sinusoidal and multifrequency forcing}

\begin{figure}[tb]
\includegraphics{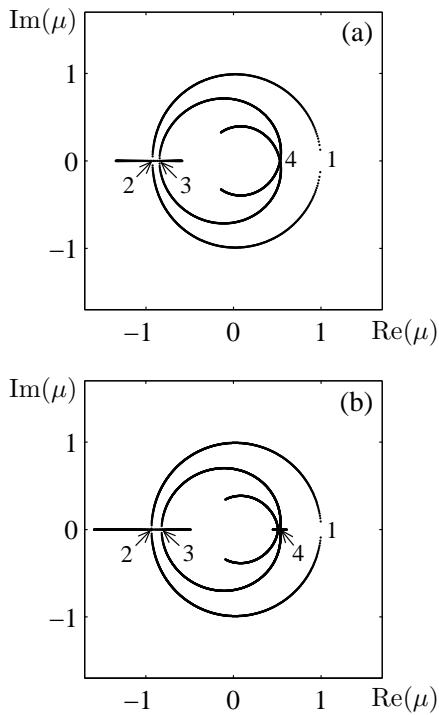}
\caption{Floquet multipliers $\mu_\pm$ of Eq.~(\ref{eq:fm}) plotted in
the complex plane over the range $k\in(0,3]$ with $f =f_c(1+1.4)$ 
(this forcing is well above the onset value for the first resonance tongue 
but below onset for the second) for (a) equally-spaced impulses, and (b) 
sinusoidal forcing.  These cases differ in that only for sinusoidal forcing 
do the Floquet multipliers split on the positive real axis. Fluid parameters 
are as in Fig.~\ref{fig:linear}. Numbers denote points described in the text.
\label{fig:fm}}
\end{figure}

We now compare the linear stability results for equally-spaced impulsive 
forcing ({\it i.e.}, $\Delta=0$) with the corresponding results for 
sinusoidal forcing
\begin{equation}
G_{\sin}(t)=f_{\sin}\sin(t).
\end{equation}
Here $f_{\sin}\equiv\frac{g_{\sin}k_0}{\omega^2}$, where $g_{\sin}$ is
the maximum (dimensioned) acceleration. (See Fig.~\ref{fig:NscTruncEq}(a) 
for an example of a neutral curve in the case of sinusoidal forcing.)  
As noted earlier, the most striking difference between these two cases, 
manifest in the neutral stability curve~(\ref{eq:asymlin}), is the lack 
of harmonic tongues for the impulsive case.  This difference is a consequence 
of how the Floquet multipliers move in the complex plane as the perturbation 
wavenumber $k$ is varied, as illustrated in Fig.~\ref{fig:fm}.  In both the
sinusoidal and the impulsive cases, the Floquet multipliers at $k=0$ (labeled 
as point 1 in Figs.~\ref{fig:fm}(a,b)) are degenerate and take the value $+1$.  
As $k$ is increased from $0$, these split into a complex conjugate pair lying 
inside the unit circle which, together, sweep out a path encircling 
the origin.  When $k$ reaches a certain value the Floquet multipliers 
meet again on the negative real axis and, at this point (labeled 2), 
are degenerate.  With additional increase in $k$, they separate and
move in opposite directions along the (negative) real axis.  If the 
magnitude of $f$ exceeds the critical value (as it does in 
Figs.~\ref{fig:fm}(a,b)), then the most negative Floquet multiplier 
eventually crosses $-1$, indicating a subharmonic instability.  In any 
case, as $k$ is increased further, these (real) Floquet multipliers 
soon reverse direction and recombine (point 3) before splitting into a 
complex conjugate pair that moves to the right, again encircling the 
origin, before merging on the positive real axis (point 4).  It
is here that the difference between the two cases emerges.
With sinusoidal forcing the Floquet multipliers split and move along
the real axis; a harmonic instability will therefore develop if the
forcing is strong enough to drive the rightmost Floquet multiplier outside 
the unit circle.  In contrast, the Floquet multipliers in the
impulsive case do {\it not} split along the positive real axis but
simply continue their trajectory around the origin as a complex
conjugate pair.

The above description follows directly from an examination of the
Floquet multipliers in (\ref{eq:MTwo}):
\begin{equation}
\mu_{\pm} = e^{-2\pi \gamma k^2}\left(\alpha\pm\sqrt{\alpha^2-1}\right),
\label{eq:fm}\end{equation}
where, for $\Delta=0$,
\begin{equation}
\alpha\equiv{1\over 2}{\rm Tr}(M_k)=
\cos(2\pi\omega_k)-\frac{1}{2}\left(\frac{fk\sin(\pi\omega_k)}
{\omega_k}\right)^2 \le 1.
\end{equation}
 From (\ref{eq:fm}) we see that the Floquet multipliers are real and distinct 
whenever $\alpha^2>1$. However, since $\alpha\le 1$, this can happen only when 
$\alpha < -1$, in which case they are negative.

\begin{figure}[tb]
\includegraphics{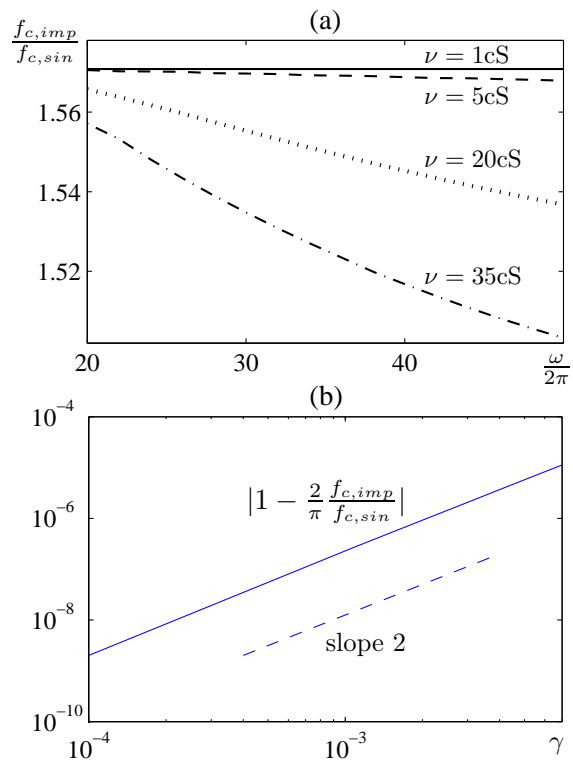}
\caption{(a) Ratio of critical forcing strength for impulsive and
sinusoidal forcing as a function of forcing frequency. Viscosities
are labeled in the figure; other parameters are as in Fig.~\ref{fig:linear}. 
The solid line for $\nu=1 {\rm cS}$ is indistinguishable from a horizontal line
at $\frac{\pi}{2}$. (b) Log-log plot of 
$|1-\frac{2}{\pi}\frac{f_{c,imp}}{f_{c,sin}}|$ as a function of $\gamma$
with $\Gamma_0=0.13$ (solid line).  For comparison, a line of slope two
is drawn nearby (dashed line).
\label{fig:sinimpcomp}}
\end{figure}

To further compare equally-spaced impulses and sinusoidal forcing, we 
consider the respective critical forcing strengths at the onset of standing waves, 
$f_{c,imp}$ and $f_{c,sin}$.  In the limit of weak damping we may expand 
(\ref{eq:asymlin}) about $\gamma=0$ and $\omega_k=\frac{1}{2}$ (near the 
minimum of the neutral stability curves for small $\gamma$)
to obtain
\begin{equation}
f_{c,imp}=\pi k_c \gamma(1+O(\gamma^2)).
\end{equation}
For sinusoidal forcing and small damping we have
\begin{equation}
f_{c,sin}=2k_c\gamma(1+O(\gamma^2)),
\end{equation}
(see, for example,~\cite{PS04}).  Thus $\frac{f_{c,imp}}{f_{c,sin}} =
\frac{\pi}{2}+O(\gamma^2)$.  This is the same ratio that arises when
considering (the first term in) the Fourier series expansion of the
two-term impulsive forcing function with $\Delta=0$:
\begin{equation}
G_{imp}(t)=\frac{2}{\pi}f_{imp}\sum_{j=0}^{\infty}
\cos((2j+1)t)\label{eq:tfs_equal}.
\end{equation}
Here we have used (\ref{eq:impacc}) and (\ref{eq:deltadef2}) with 
$f_n=(-1)^nf_{imp}$.  In Fig.~\ref{fig:sinimpcomp}(a) we plot 
the ratio $\frac{f_{c,imp}}{f_{c,sin}}$ as a function of forcing
frequency (the most easily tuned parameter in experiments) for several
viscosities.  As anticipated, for small $\nu$ this ratio is well
approximated by $\frac{\pi}{2}$.  Furthermore, the ratio decreases
with increasing viscosity (equivalently, $\gamma$) with a deviation
$|1-\frac{2 f_{c,imp}}{\pi f_{c,sin}}| = O(\gamma^2)$ (see
Fig.~\ref{fig:sinimpcomp}(b)).  

We may also consider the effect of adding a second term, $\cos(3t)$,  
to the truncation of the Fourier series~(\ref{eq:tfs_equal}) and 
examine $f_{c,2}$, the critical value of this two-frequency forcing
function, in the weak damping limit.  (In general, $f_{c,M}$ denotes
the dimensionless forcing strength at onset for the $M$-term
truncation of (\ref{eq:tfs_equal}).)  It is demonstrated in~\cite{TS02} 
that for a two-frequency forcing function with commensurate frequencies
$m\omega$ and $n\omega$ ($m\omega$ is assumed to drive the primary 
instability), the $n\omega$ frequency component is {\it destabilizing}
when $\frac{m}{n}<\sqrt[4]{2}\approx 1.19$.  By this we mean that the
threshold for instability is lower in the two-frequency case than in
the single frequency case.  For the two-term truncation
of~(\ref{eq:tfs_equal}), $\frac{m}{n}=\frac{1}{3}$ and hence we expect
that $f_{c,2}<f_{c,1}=\frac{\pi}{2}f_{c,sin}$, {\it i.e.},
$\frac{f_{c,2}}{f_{c,sin}}<\frac{\pi}{2}$.  Indeed, this is consistent 
with our finding that $\frac{f_{c,imp}}{f_{c,sin}} < \frac{\pi}{2}$ for 
impulsive forcing (see Fig.~\ref{fig:sinimpcomp}(a)).

\begin{figure}[tb]
\includegraphics{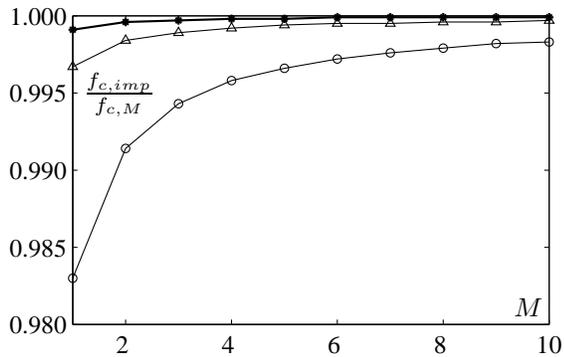}
\caption{Ratio of critical forcing strengths for equally-spaced
impulses and the $M$-term truncated Fourier series, as a function
of $M$, for $\Gamma_0=0.04$ and $\gamma=0.02$ (filled circles),
$\gamma=0.04$ (triangles), $\gamma=0.1$ (open circles).
\label{fig:LinTrunc}}
\end{figure}

As expected, adding more terms to the truncation of the
Fourier series~(\ref{eq:tfs_equal}) results in a closer approximation 
to the critical forcing strength $f_{c,imp}$, as shown 
in Fig.~\ref{fig:LinTrunc}.  Despite this good agreement between
$f_{c,imp}$ and $f_{c,M}$, the $M$-term truncated Fourier series is
far less successful in capturing the behavior of the resonance tongues
away from the minimum of the first tongue; this can be seen in
Fig.~\ref{fig:NscTruncEq}.  For $\gamma=0.02$ and $\Gamma_0=0.04$, for
example, $f_{c,M}$ approximates $f_{c,imp}$ remarkably well even for
small $M$ (filled circles in Fig.~\ref{fig:LinTrunc}); however, with
$M=50$, only the first four resonance tongues plausibly resemble those
for impulsive forcing (see Fig.~\ref{fig:overlay}).

\section{Weakly nonlinear analysis}
\label{sec:nonlinear}

\subsection{Weakly nonlinear calculation}

We now extend our analysis of $N=2$ impulses to the weakly
nonlinear regime in the case of one-dimensional waves.

Following~\cite{SS99}, we perform a multi-scale expansion:
\begin{subequations}
\begin{align}
h(x,t,T) & =  \epsilon h_1(x,t,T)+\epsilon^2 h_2(x,t,T) +
\cdots, \\
\Phi(x,t,T) & =  \epsilon \Phi_1(x,t,T)+\epsilon^2 \Phi_2(x,t,T)
+ \cdots, 
\end{align} 
\end{subequations}
where $\epsilon \ll 1$, the slow time is $T = \epsilon^2 t$, and the 
forcing amplitude is $f = f_c(1 +\epsilon^2 f_2)$. We seek 
spatially-periodic solutions in separable Floquet-Fourier form:
\begin{subequations}
\begin{align}
h_{1}(x,t,T) & =  Z(T) p_1(t) e^{i k_c x}+ c.c., \\
h_{2}(x,t,T) & =  Z^2(T) p_{2}(t) e^{2i  k_c x}+c.c., \\
\Phi_{1}(x,t,T) & =  Z(T) q_1(t) e^{i k_c x} + c.c., \label{eq:phi1} \\
\Phi_{2}(x,t,T) & =  Z^2(T) q_2(t) e^{2i  k_c x}+ c.c.,
\end{align}
\end{subequations}
where the critical wave number $k_c$, and the periodic eigenfunction
$p_1(t)$ are determined from the linear stability analysis.
Specifically, we take $(k_c,f_c)$ to be the minimum of the neutral
stability curves and use (\ref{eq:p}), evaluated at $k=k_c$ and with 
$|A_{k_c,n}|=1$, to obtain $p_1(t)$.  The phase of $A_{k_c,n}$ is determined  
by the map~(\ref{eq:MTwo}), evaluated at $(k,f)=(k_c,f_c)$, and the 
(subharmonic) condition $p_1(t+2\pi)=-p_1(t)$ (equivalently, 
$A_{k_c,n+2}=-A_{k_c,n}$):    
\begin{align}
\frac{A_{k_c,n}^i}{A_{k_c,n}^r} = \frac{\cos(2\pi\omega_{k_c})-F_nc_ns_{n+1}
+e^{2\pi\gamma k_c^2}}
{\sin(2\pi\omega_{k_c})-F_ns_ns_{n+1}}.
\label{eq:Amagphase}
\end{align}
This expression defines the critical eigenmode associated with
the instability.   The subharmonic assumption $A_{k_c,n+2}=-A_{k_c,n}$
is valid when the instability is associated with the first resonance 
tongue, as it is for all parameters we have investigated.  
Plugging (\ref{eq:phi1}) into (\ref{eq:zva}) yields
$q_1(t) = \frac{1}{k_c}(p_1^\prime(t)+\gamma k_c^2 p_1(t))$ at leading
order in $\epsilon$.  Since in this section $k$ will generally be fixed 
at its critical value, we hereafter drop this subscript from 
many expressions ({\it e.g.}, we write $A_n$ for $A_{k_c,n}$) or 
replace it with a subscript 1 (such as with $p_1(t)$).

At second order we find the equation ({\it cf.}~(\ref{eq:peq}))
\begin{align}
q_1^2 = &\ p_2^{\prime\prime}+2\gamma (2k_c)^2p_2^\prime \nonumber \\ 
&+[\gamma^2 (2k_c)^4 +\Gamma_0 (2k_c)^3 +(G_0 - G(t))2k_c]p_2, 
\label{eq:p2}
\end{align} 
which has the solution
\begin{align}
p_2(t) =  &\ aA_{n}^2 e^{2(-\gamma k_c^2 + i \omega_{1})(t -
t_{n})}+ \frac{1}{2}b|A_n|^2 e^{-2\gamma k_c^2 (t-t_n)} \nonumber \\ 
&+ B_n e^{(-\gamma(2k_c)^2 + i \omega_{2})(t - t_{n})} + c.c., 
\end{align}
for $t \in (t_n, t_{n+1})$.  Here
\begin {equation}
a = \frac{-2k_c\omega_{1}^2}{\omega_{2}^2-4\omega_{1}^2+4\gamma^2
k_c^2+8i\gamma k^2 \omega_{1}}, \label{eq:a}
\end{equation}
\begin{equation}
b = \frac{4k_c\omega_{1}^2}{\omega_2^2+4\gamma^2 k_c^4},
\label{eq:b}
\end{equation}
and $\omega_{1}\ (\omega_{2})$ is the frequency of a wave with 
wavenumber $k_c\ (2k_c)$ obtained from the dispersion relation (\ref{eq:disp}).
Note that the ``homogeneous" solution to (\ref{eq:p2}) must be included
({\it i.e.}, $B_n \neq 0$) to ensure that $p_2(t)$ is continuous.  Using
this continuity condition and the jump condition on $p^{\prime}_2$
(obtained as in the linear stability calculation), we relate $B_{n+1}$
to $B_n$ through the map
\begin{equation} \left( \begin{array}{c}
    B_{n+1}^r \\ B_{n+1}^i
\end{array} \right)
= e^{-4\gamma k_c^2 d_n}
    M_{2,n}
\left( \begin{array}{c}
    B_n^r \\ B_n^i
\end{array} \right)
+R_n+bS_n. \label{eq:B}
\end{equation}
Here $M_{2,n}$ is the matrix $M_{k,n}$ of (\ref{eq:Mn}) with $k=2k_c$;
the vectors $R_n$ and $S_n$ are given in the Appendix.  To obtain
$B_n$, we require that $B_{n+2}=B_n$. (This harmonic condition is due to 
the fact that $q_1^2$, the driving term in (\ref{eq:p2}), is $2\pi$-periodic.)  

At third order we obtain the solvability condition (see~\cite{SS99})
\begin{equation}
\tau\frac{dZ}{dT} = L f_2 Z + (C_{res}+C_{non})|Z|^2Z,
\label{eq:cubic}\end{equation}
where $\tau$, $L$, $C_{res}$ and $C_{non}$ are given by
\begin{subequations}\label{eq:O3expressions}
\begin{align}
\tau & =  \frac{1}{2\pi}\int_{0^-}^{4\pi^-}(p_1^\prime+\gamma k_c^2 p_1)\tilde{p}_1dt\\
& =  \frac{1}{2\pi}(A_0\tilde{A_0}(e^{i\omega_1d_0}-1-i\omega_1) \nonumber \\
&  \ \ \ \ \ \  +A_1\tilde{A_1}(e^{i\omega_1 d_1}-1-i\omega_1))+c.c., \nonumber \\
L & =  \frac{k_c}{4\pi}\int_{0^-}^{4\pi^-}
{G_{imp}(t)\over f} p_1\tilde{p}_{1}dt \\
& = 
\frac{k_c}{2\pi}(A_0\tilde{A_0}+A_0\overline{\tilde{A}}_0-A_1
\tilde{A_1}-A_1\overline{\tilde{A}}_1)
+ c.c.,\nonumber \\
C_{res} & =  -\frac{k_c^2}{2\pi}\int_{0^-}^{4\pi^-}
[(q_{1}p_{2})^\prime+\gamma k_c^2
q_{1}p_{2}]\tilde{p}_{1}dt, \\ 
C_{non} & =  \frac{k_c^3}{4\pi}\int_{0^-}^{4\pi^-}
[-(p_{1}^2q_{1})^\prime-\gamma
k_c^2p_{1}^2q_{1}+k_cq_{1}^2p_{1} \\
&  \ \ \ \ \ \  +\frac{3}{2}k_c^2\Gamma_0p_{1}^3]\tilde{p}_{1}dt.\nonumber
\end{align}
\end{subequations}
The full expressions for $C_{res}$ and $C_{non}$ are provided in the Appendix. 
In equations~(\ref{eq:O3expressions}) we use $\tilde{p}_{1}$ to indicate   
the solution of the adjoint linear problem, which has the same form as 
(\ref{eq:pdiff}) but with $\gamma \rightarrow -\gamma$.  Hence
\begin{equation} 
\tilde{p}_{1}(t)=\tilde{A}_{n}e^{(\gamma k^2+i\omega_1)(t-t_n)}+c.c., 
\quad t\in(t_n,t_{n+1})\ ,
\label{eq:adjoint}
\end{equation} 
with the complex conjugate of $\tilde{A}_n$ denoted by 
$\overline{\tilde{A}}_n$.
The map relating $\tilde{A}_{n+1}$ to $\tilde{A}_n$ is similarly obtained 
from (\ref{eq:A}) by replacing $\gamma$ with $-\gamma$ and setting $k=k_c$.  
Note that in (\ref{eq:cubic}) we separate the cubic coefficient $C$ into 
two distinct contributions, $C_{res}$ (resonant) and $C_{non}$ (nonresonant).  
The $C_{res}$ term can be traced to quadratic nonlinearities in (\ref{eq:zv})
({\it i.e.}, terms in which the $k_c$ and $2k_c$ modes interact), while
$C_{non}$ derives from cubic nonlinearities and involves only the
critical $k_c$ mode.

\subsection{Weakly nonlinear results}

\begin{figure}[tb]
\includegraphics{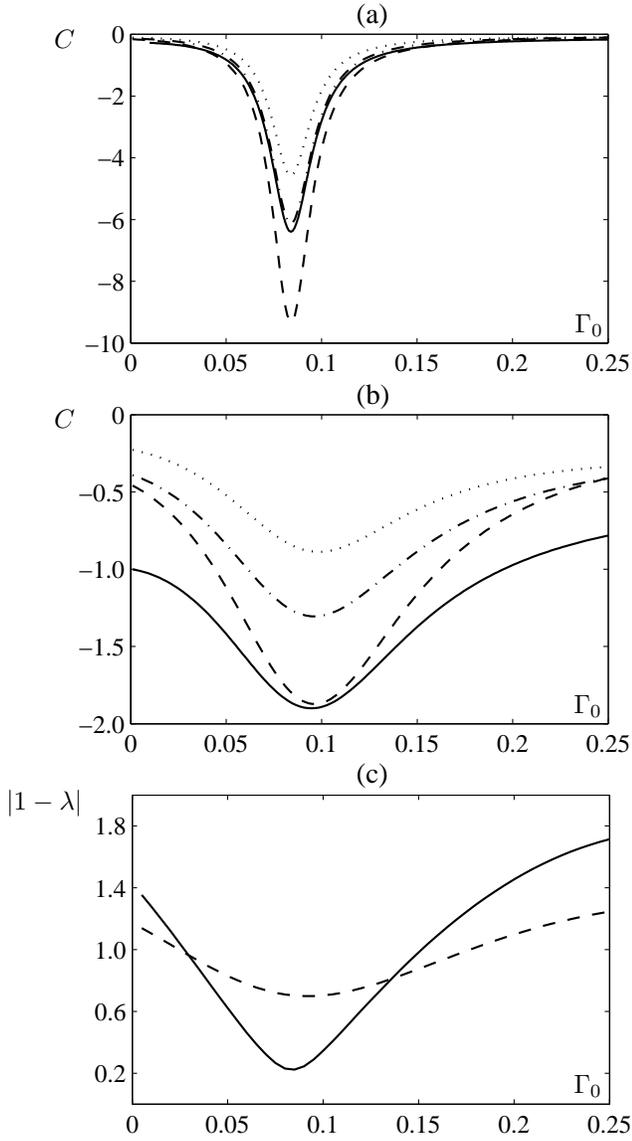}
\caption{Cubic coefficient in (\ref{eq:cubic}) as a function of
$\Gamma_0$ with (a) $\gamma = 0.01$ and (b) $\gamma=0.05$. The dashed line
indicates the result for impulsive forcing with $\Delta=0.5$; the dot-dashed
line is for $\Delta=0$; the dotted is for $\Delta=-0.5$; the solid line
is for sinusoidal forcing. (c) $|1-\lambda|$ as a function of $\Gamma_0$
where $\lambda$ and $\overline{\lambda}$ are the Floquet 
multipliers~(\ref{eq:fm})  of the
$2k_c$ mode for $\Delta=0$ and  $\gamma = 0.01$
(solid line) or $\gamma = 0.05$ (dashed line).}
\label{fig:cubic}
\end{figure}

We now examine in greater detail the cubic coefficient
$C=C_{res}+C_{non}$, which determines the nonlinear saturation of the
instability to standing waves.  Figs.~\ref{fig:cubic}(a,b) show $C$ 
as a function of the capillary parameter $\Gamma_0$ for various choices 
of $\Delta$.  We find, for most parameters, that $C(\Gamma_0)<0$, ensuring 
that the bifurcation to standing waves is supercritical.  (We find 
subcritical bifurcations only in the capillary wave regime, 
$\Gamma_0 \approx 0.25$, and with extreme asymmetry, {\it e.g.} 
$\Delta\approx 0.99$.)

A dominant feature of the curves shown in Figs.~\ref{fig:cubic}(a,b)
is the dip in $C$ that reaches a minimum value at
$\Gamma_0=\Gamma_{res} \sim 0.09$. This resonance feature is apparent
both with impulsive forcing (non-solid lines in
Figs.~\ref{fig:cubic}(a,b)) and with sinusoidal forcing (solid lines).
For this figure we have used a normalization convention for $p_1(t)$
in the sinusoidal case which agrees with our normalization convention
in the impulsive case ($|A_{k_c,n}|=1$) in the limit $\gamma\to 0$.
Note that the sinusoidal and impulsive results agree quantitatively for
$\gamma=0.01$ and $\Delta=0$. For larger damping ($\gamma=0.05$), the
sinusoidal curve is shifted relative to the impulsive ones indicating
a difference between their nonresonant contributions $C_{non}$ to
$C$.

In the sinusoidal case, the dip in $C$ is easily attributed to a 1:2
spatio-temporal resonance, {\it i.e.}, at $\Gamma_{res}$ we have
$2\omega_{1} = \omega_{2}$ (see~\cite{ZV97}, for example).  The
strength of this feature, which scales with $1/\gamma$, is due to
the strong coupling between the first subharmonic and the first
harmonic modes, which become spatially commensurate at $\Gamma_{res}$.
Since harmonic resonance tongues are entirely absent from the neutral
stability picture for equally-spaced impulses, it is perhaps
surprising that the dip appears in this case as well.  However, we can
understand its origin in the following way (see~\cite{SS99} for an
analogous argument in the case of two-frequency forced Faraday waves).
We express the height of spatially-periodic surface waves at integer
multiples of the forcing period in the form
\begin{equation}
h(x,t_m)=z_{m,\mu}e^{ik_cx}+w_{m,\lambda}e^{i2k_cx}
+v_{m,\overline{\lambda}}e^{i2k_cx}+\cdots+c.c.,
\label{eq:hmap}
\end{equation}
where $t_m=2\pi m$, $m\in\mathbb{Z}$.  Here $\mu$ is the (real) Floquet
multiplier of the $k_c$ mode while $\lambda$ and $\overline{\lambda}$
are the Floquet multipliers of the $2k_c$ mode(s), assumed to be
complex.  From the linear stability analysis, we know that the
critical standing waves are subharmonic, {\it i.e.}, at onset
$\mu=-1$.  Considering the symmetries of the problem, including
spatial translation and reflection, we arrive at a stroboscopic map
from the $m^{th}$ to $(m+1)^{st}$ period:
\begin{eqnarray}
z_{m+1} & = & -z_m+(q_1w_m+\overline{q}_1v_m)\overline{z}_m+a_1|z_m|^2z_m+\cdots, \nonumber \\
w_{m+1} & = & \lambda w_m+q_2z_m^2 +\cdots, \\
v_{m+1} & = & \overline{\lambda} v_m+\overline{q}_2z_m^2+\cdots, \nonumber
\end{eqnarray}
where we have kept the terms that contribute at leading order to the cubic
coefficient in (\ref{eq:cubic}).  Because $w_m$ and $v_m$ are damped
modes ($|\lambda|<1$), we may perform a center manifold reduction to
eliminate them.  This gives, at leading order,
\begin{align}
w_m = -\frac{q_2}{(1-\lambda)}z_m^2, \quad
v_m = -\frac{\overline{q}_2}{(1-\overline{\lambda})}z_m^2,
\end{align}
resulting in the map
\begin{equation}
z_{m+1}=-z_m+\left(a_1-\frac{q_1q_2}{(1-\lambda)}-
\frac{\overline{q}_1\overline{q}_2}
{(1-\overline{\lambda})}\right)|z_m|^2z_m.
\label{eq:map}
\end{equation}
If $\lambda$ and $\overline{\lambda}$ are close enough to one, then
the slaved modes can make a large contribution to the cubic
coefficient $C$.  Using the Floquet analysis of Section~\ref{ssec:linstabanal},
we confirm that for $\Gamma_0=\Gamma_{res}$ the Floquet multipliers
associated with the $2k_c$ mode come closest to one, {\it i.e.}, at
that point the Floquet multipliers cross the positive real axis (see
Fig.~\ref{fig:cubic}(c)).  This resonant contribution to the cubic
coefficient $C$ becomes increasingly pronounced as the damping decreases 
(compare the vertical scales in Figs.~\ref{fig:cubic}(a) and (b)).

\begin{figure}[tb]
\includegraphics{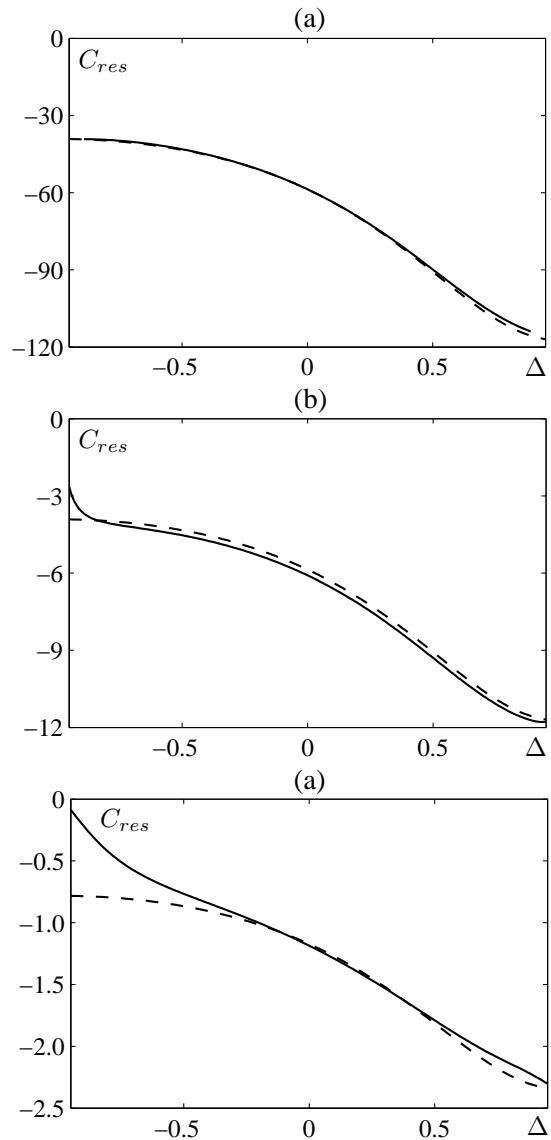}
\caption{Resonant contribution $C_{res}$ to the cubic coefficient of
Eq.~(\ref{eq:cubic}) as a function of the asymmetry parameter $\Delta$
over the range $[-0.95,0.95]$, calculated for the Zhang-Vi\~{n}als 
equations using~(\ref{eq:cres}) (solid line) and using multifrequency 
forcing results~(\ref{eq:Cres2f}) with $\alpha_1=0.235$ (dashed line). 
Plots are for (a) $\gamma=0.001$, (b) $\gamma=0.01$, and (c) $\gamma=0.05$.
\label{fig:cubicofasym}}
\end{figure}

A comparison of the (non-solid) curves in Figs.~\ref{fig:cubic}(a,b)
reveals that the spacing of the impulses can have a significant effect 
on the magnitude of the resonance feature.  Some explanation for this 
behavior is provided by recent results~\cite{PTS04,TPS04} on 
multifrequency forced Faraday waves in which the form of the cubic 
amplitude equations is obtained from the spatial symmetries and 
the (weakly broken) symmetries of time translation, time reversal, and 
Hamiltonian structure.  Such a procedure yields important qualitative 
features of the behavior of the critical modes and other damped modes 
driven through nonlinear interactions, including a prediction about 
which particular damped modes are most important and which forcing 
frequencies (and corresponding temporal phases) are most effective in 
enhancing or otherwise controlling their effect.  Here we are interested 
in the Fourier series expansion for an impulsive forcing function with 
two impulses per period:
\begin{equation}
G_{imp}(t)=\frac{f_{imp}}{2\pi}\sum_{j=1}^{\infty}
(1-e^{-ij((1+\Delta)\pi)})e^{ijt} + c.c.
\label{eq:truncfs}
\end{equation} 
In \cite{PTS04,TPS04} the contribution of the damped $2k_c$ mode to the 1:2 
spatio-temporal resonance is found to be most affected by the forcing 
frequency $2\omega$ (where $\omega$ is the primary driving frequency) and 
its phase.  This result suggests that we focus on the drastically truncated 
two-term Fourier series
\begin{equation}
G_2(t)=\frac{f_{imp}}{2\pi}\left[(1+e^{-i\pi\Delta}) e^{it}+
(1-e^{-2i\pi\Delta})e^{2it}+c.c.\right]. \label{eq:twotrunc}
\end{equation}
At leading order in the damping $\gamma$, the onset of standing waves 
occurs when the magnitude of the first term in (\ref{eq:twotrunc})  
becomes equal to $\gamma$ (see, for example,~\cite{PS04}).  Using 
\begin{equation}
\frac{f_{imp}}{2\pi}|1+e^{-i\pi\Delta}|=\gamma ,
\end{equation}
and simplifying the result through an appropriate time 
translation, we may rewrite (\ref{eq:twotrunc}) at onset as
\begin{equation}
G_{2}(t)=\gamma e^{it}+F_2e^{2it}+c.c.,
\end{equation}
where
\begin{equation}
F_2 = 2i\gamma\sin\big(\Delta\frac{\pi}{2}\big).
\end{equation}
 From Table 1 in~\cite{PTS04} we find the predicted magnitude, at 
leading order in $\gamma$, of the resonant dip in the cubic 
coefficient at $\Gamma_{res}$,
\begin{equation}
C_{res}^{2f}=-\alpha_1\frac{|L_3|+\mu_i|f_{2\Omega}
|\sin\Phi}{|L_3|^2-|\mu_if_{2\Omega}|^2},
\end{equation}
Here $|f_{2\Omega}|=|F_2|=2\gamma\sin\left(\Delta\frac{\pi}{2}\right)$, 
$\Phi=\arg(F_2)={\rm sign}(\Delta)\frac{\pi}{2}$, $L_3$ is the linear 
damping coefficient of the $2k_c$ mode ($L_3=(2k_c)^2\gamma\approx4\gamma$), 
and $\mu_i$ is the coefficient of the linear parametric driving term for 
the $2k_c$ mode ($\mu_i=\frac{2k_c}{2}\approx 1$).  
Thus $C_{res}^{2f}$, as a function of the asymmetry parameter $\Delta$, becomes
\begin{equation}
C_{res}^{2f}(\Delta) \approx -\frac{\alpha_1}{2 \gamma}\ \frac{1}
{2-\sin\left(\frac{\Delta\pi}{2}\right)} \,.
\label{eq:Cres2f}
\end{equation}
The constant $\alpha_1$ is positive and must be determined by a
nonlinear calculation; it depends on $\Gamma_0$, but not, at leading
order, on $\gamma$.  Here we choose $\alpha_1$ such that the values of
the multifrequency prediction $C_{res}^{2f}(\Delta)$ and our direct
calculation of $C_{res}(\Delta)$ at $\Gamma_0=\Gamma_{res}$ from
(\ref{eq:cres}) agree for small damping ($\gamma=0.001$) at
$\Delta=0$.  (We then use the same estimate of $\alpha_1$ throughout this
comparison.)  We compare $C_{res}(\Delta)$ with the multifrequency
prediction in Fig.~\ref{fig:cubicofasym} and find good agreement,
especially for small $\gamma$ where the multifrequency results are
expected to be valid.  This agreement breaks down as $|\Delta|$
approaches unity (even more so for larger $\gamma$), an effect that is
not overly surprising given that $|\Delta|\to 1$ is an unphysical
limit requiring an infinite value of $f_{imp}$ to produce standing
waves.  Note that $C_{res}^{2f}$, given by~(\ref{eq:Cres2f}), scales with 
$\gamma^{-1}$, which explains the difference in scales evident in 
Fig.~\ref{fig:cubic}(a,b).

\section{Discussion and Conclusions}

In this paper we examined the problem of Faraday waves parametrically 
excited by a periodic sequence of impulses (delta functions).  We showed 
how to extend the linear stability analysis presented in~\cite{BJ96} to 
include more general impulsive forcing functions, and investigated the 
weakly nonlinear regime for one-dimensional surface waves in the case of 
two impulses per period, comparing both linear and nonlinear results with 
the more established cases of sinusoidal and multifrequency forcing.  Our  
analytical and numerical work was conducted within the framework of the 
Zhang-Vi\~nals Faraday wave model, appropriate for describing small amplitude 
surface waves on deep, weakly-viscous fluid layers.  

One theoretical advantage of using an idealized impulsive forcing
function is that it allows for an exact, albeit generally implicit,
expression for the neutral stability curves associated with the
primary instability.  Moreover, since the stroboscopic map
characterizing the linear problem can be explicitly constructed,
Floquet multipliers can be readily determined as a function of forcing
and fluid parameters.  This stands in sharp contrast to the sinusoidal
and multifrequency cases where even the neutral curves must be
determined numerically or through an asymptotic expansion that assumes
weak damping and forcing. A further consequence of exactly solving the 
linear problem is that we are then able to derive explicit, analytic 
expressions for the amplitude of weakly nonlinear, one-dimensional 
surface waves as a function of forcing and fluid parameters.

As noted by~\cite{BJ96}, in the simplest case of $N=2$ impulses per period,
harmonic instabilities from the flat fluid state cannot be excited with 
equally-spaced impulses.  Despite this fact, we found, through a weakly 
nonlinear analysis, that the 1:2 spatio-temporal resonance remains the 
dominant feature in typical parameter regimes.  Symmetry arguments of the 
type used in~\cite{SS99} showed that, as in the case of sinusoidal forcing, 
harmonic modes can be driven nonlinearly by the critical modes --- hence the 
resonance effect persists.

By varying the spacing between the up and down impulses making up the $2\pi$-periodic 
forcing function for $N=2$, we found that the magnitude of the 1:2 resonance effect 
depends dramatically, and monotonically, on the corresponding asymmetry parameter 
$\Delta$.  Appealing to recent results from multifrequency forcing theory~\cite{TPS04}, 
valid in the limit of weak damping, we obtained a prediction of this dependence 
based on a truncated two-term Fourier series approximation.   This prediction 
agreed quite well with the results for impulsive forcing despite the severity of the 
approximation involved.

We envision several ways in which this work could be extended.  For example, 
the methods used in this paper could be easily applied to the case of piecewise 
constant forcing, a type of driving function that has been explored in other systems 
(see \cite{HC73,Y78}, for instance) where it also led to analytic results.  
It would further be of interest to extend our weakly nonlinear analysis to the case of 
two-dimensional patterns.  In such a context, additional questions of pattern 
selection, as well as more specific comparisons with multifrequency forcing, could 
be explored.  One could try, for example, to construct impulsive forcing functions 
that mimic specific effects seen with multifrequency forcing.  In this way
the analytic results available for impulsive forcing would complement
the numerical~\cite{BF95,BET96,TS02} and experimental 
results~\cite{EF93,M93,EF94,KPG98,AF98,AF00a,AF02,EF04} reported with multifrequency 
forcing.

\acknowledgments

We thank Yu Ding, Cristian Huepe, Chad Topaz, and Paul Umbanhowar for
many helpful discussions.  A.C. is grateful for fellowship support
through NSF-IGERT grant DGE-9987577.  M.S. acknowledges support
through NSF Grant DMS-0309667.  The research of M.S. and J.P. was
supported in part by NASA Grant NAG3-2364.  J.P. further acknowledges
support through EPSRC grant GR/R52879/01.

\appendix
\section{Analytic Expressions}

The vectors given in (\ref{eq:B}) are
\begin{widetext}
\begin{equation} R_n = \left(
\begin{array}{c}
    Re[a(A_n^2e^{(-\gamma k_c^2 +i\omega_1)2d_n}-A_{n+1}^2)]\\
    \frac{2}{\omega_2}Re[a(A_n^2(e^{(-\gamma k_c^2
    +i\omega_1)2d_n}((-1)^nk_cf_c-\gamma
    k_c^2-i\omega_1)+A_{n+1}^2 (\gamma k_c^2+i\omega_1)))]
\end{array} \right),
\end{equation}
and
\begin{equation} S_n = \left(
\begin{array}{c}
    \frac{1}{2}(|A_n|^2e^{-2\gamma k_c^2 d_n} -|A_{n+1}|^2)\\
    \frac{1}{\omega_2}(|A_n|^2e^{-2\gamma k_c^2 d_n}((-1)^n2k_cf_c-\gamma
    k_c^2)+|A_{n+1}|^2\gamma k_c^2)
    \end{array} \right)
.\end{equation}
\end{widetext}
The expressions for $a$ and $b$ in these vectors
are given by (\ref{eq:a}) and (\ref{eq:b}), respectively, and
$\omega_1$ ($\omega_2$) satisfies the dispersion relation (\ref{eq:disp})
with wavenumber $k_c$ ($2k_c$).

The full expressions for the components ($C_{res}$ and $C_{non}$) of the 
cubic coefficient in the solvability condition (\ref{eq:cubic}) are 
\begin{widetext}
\begin{eqnarray}
C_{res} & = & \frac{-2i\omega_1k_c}{\pi}\sum_{j=0,1}
A_jB_j\tilde{A_j}\frac{\beta_1^+-i\omega_1}{\beta_1^+}
(e^{\beta_1^+d_j}-1)-\bar{A_j}B_j\bar{\tilde{A_j}}
\frac{\beta_1^-+i\omega_1}{\beta_1^-}(e^{\beta_1^-d_j}-1)
\label{eq:cres}\\
& & +\left(A_jB_j\bar{\tilde{A_j}}\frac{\beta_2+i\omega_1}{\beta_2}
-\bar{A_j}B_j\tilde{A_j}\frac{\beta_2-i\omega_1}{\beta_2}\right)
(e^{\beta_2d_j}-1)\nonumber\\
& & +\left(aA_j^3\bar{\tilde{A_j}}\frac{\beta_3+i\omega_1}{\beta_3}
+(b-a)|A_j|^2A_j\tilde{A_j}\frac{\beta_3-i\omega_1}{\beta_3}\right)
(e^{\beta_3d_j}-1)\nonumber\\
& & +aA_j^3\tilde{A_j}\frac{\beta_4-i\omega_1}{\beta_4}(e^{\beta_4d_j}-1)
+(b-a)|A_j|^2A_j\bar{\tilde{A_j}}\frac{2\gamma k_c^2-i\omega_1}
{2\gamma k_c^2}(e^{-2\gamma k_c^2d_j}-1) + c.c.,\nonumber\\
C_{non} & = & \sum_{j=0,1}\left(A_j^3\bar{\tilde{A_j}}+|A_j|^2A_j\tilde{A_j}\right)
\frac{3\Gamma_0k_c^5-i\omega_1k_c^2\beta_3}{\pi\beta_3}(e^{\beta_3d_j}-1)\\
& & +A_j^3\tilde{A_j}\frac{3\Gamma_0k_c^5-2i\omega_1k_c^2\beta_4}{4\pi\beta_4}
(e^{\beta_4d_j}-1)+|A_j|^2A_j\bar{\tilde{A_j}}
\frac{4i\gamma \omega_1k_c^4-9\Gamma_0k_c^5}{8\pi\gamma k_c^2}
(e^{-2\gamma k_c^2d_j}-1)+c.c.,\nonumber
\end{eqnarray}
where 
\begin{equation}
\beta_1^{\pm}=-4\gamma k_c^2+i(\pm2\omega_1+\omega_2),\ 
\beta_2=-4\gamma k_c^2+i\omega_2,\ 
\beta_3=-2\gamma k_c^2+2i\omega_1,\ 
\beta_4=-2\gamma k_c^2+4i\omega_1.
\end{equation}
\end{widetext}

\bibliography{catllaetal}

\end{document}